\documentclass[11pt]{article}
\usepackage{amsmath,amssymb,color,graphics,epsfig,cite}

\usepackage{amsfonts}

\newcommand{\be}{\begin{equation}}
\newcommand{\ee}{\end{equation}}
\newcommand{\bea}{\setlength\arraycolsep{2pt} \begin{eqnarray}}
\newcommand{\eea}{\end{eqnarray}}
\newcommand{\nn}{\nonumber}

\def\ft#1#2{{\textstyle{\frac{\scriptstyle #1}{\scriptstyle #2} } }}

\def\0{{\sst{(0)}}}
\def\1{{\sst{(1)}}}
\def\2{{\sst{(2)}}}
\def\3{{\sst{(3)}}}
\def\4{{\sst{(4)}}}
\def\5{{\sst{(5)}}}
\def\6{{\sst{(6)}}}
\def\7{{\sst{(7)}}}
\def\8{{\sst{(8)}}}
\def\sst#1{{\scriptscriptstyle #1}}

\begin{document}

\begin{center}
{\Large {\bf New Rotating Black Hole in Electromagnetic Fields: \\ Cosmological Horizon without Cosmological Constant}}

\vspace{20pt}

Liang Ma and H. L\"u

\vspace{10pt}

{\it Center for Joint Quantum Studies, Department of Physics,\\
School of Science, Tianjin University, Tianjin 300350, China }

\vspace{40pt}

\underline{ABSTRACT}
\end{center}

We obtain a new electrovacuum and related background spacetime that contains a cosmological horizon supported entirely by the electromagnetic field.  The local solution belongs to the Kundt class of type D, but is globally distinct.  We further construct an exact solution describing a Kerr black hole in this background. We study the global structure including horizons and singularities, and derive the first law of black hole thermodynamics. The emergence of a cosmological horizon in Einstein-Maxwell gravity without invoking a positive cosmological constant or dark energy is tantalizing, and may provide a new avenue for exploring cosmological and astrophysical phenomena related to black holes and the late-time cosmology.

\vfill {\footnotesize maliang0@tju.edu.cn\ \ \ mrhonglu@gmail.com}


\thispagestyle{empty}
\pagebreak



\section{Introduction}

Current cosmological observations indicate that the Universe is undergoing accelerated expansion and is well described by an asymptotically de Sitter (dS) spacetime \cite{SupernovaSearchTeam:1998fmf,SupernovaCosmologyProject:1998vns,Planck:2018vyg}. Conventional explanations invoke a positive cosmological constant or more general dark-energy components \cite{Copeland:2006wr,Frieman:2008sn}. Many dynamical dark-energy models are effectively described by exotic matter fields \cite{Caldwell:1999ew}. The characteristic feature of accelerated expansion is the presence of a cosmological horizon.
Consequently, static or stationary spacetimes possessing a cosmological horizon are generally believed to require a positive cosmological constant or other forms of dark energy. In this work, however, we demonstrate that a cosmological horizon can emerge within Einstein-Maxwell theory itself, where the horizon is supported entirely by appropriately configured electric or magnetic (E/M) fields.

Our construction builds upon the recently obtained Ricci-flat ${\cal B}$-deformed Kerr metric \cite{Ma:2026otg}, which itself arises from applying an Ernst transformation \cite{Ernst:1976mzr,Ernst:1976bsr} to the Kerr-Bertotti-Robinson (KBR) black hole constructed by Podolsk{\'y} and Ovcharenko (PO) \cite{Podolsky:2025tle,Ovcharenko:2025cpm}.
(See also \cite{Astorino:2025lih,Astorino:2026okd,Ovcharenko:2026byw,Barrientos:2026shy,
Hu:2026slp,Herdeiro:2026jem,DiPinto:2026rvp,Yu:2026tqk}.) The new rotating black hole can serve as a seed solution: by acting on the solution with the constant $SU(2,1)$ transformation matrix \(U\) introduced in \cite{Gibbons:2013yq}, which involves constant complex variables $B\pm {\rm i} E$, we can construct spacetimes immersed in external E/M fields, thereby generating a wider family of exact black hole solutions within Einstein-Maxwell theory $\mathcal{L}=R-F^2$. In particular the KBR solution emerges when ${\cal B}=B$. The structure of the global transformation suggests that we may instead take ${\cal B}={\rm i} E$. Concretely, we set the parameter of \cite{Ma:2026otg} to be ${\cal B}={\rm i} {\cal E}$, under which the metric remains real. We then perform the magnetization using the matrix $U$. We consider two particular choices, namely \(B=0,\ E=\mathcal{E}\) or \(E=0,\ B=\mathcal{E}\), which give rise to a new rotating solution describing Kerr black hole immersed in a new electrovacuum.

Setting ${\cal B}={\rm i} {\cal E}$ has the surprising effect that a cosmological horizon emerges whose radius is inversely proportional to ${\cal E}$. The spacetime background is thus enclosed by the cosmological horizon supported entirely by the E/M field, beyond which the metric is cosmological. The paper is organized as follows. In Section 2, we present the local solution, its static limit and the associated electrovacuum background. In Section 3, we present a brief analysis of the global structure, focusing on the static configuration. We present representative plots of E/M fields in black hole spacetimes that involve event, cosmological horizons and spacetime singularities. In Section 4, we derive the first law of black hole thermodynamics for the most general rotating solution. We conclude the paper in Section 5. We present comparative graphs of Bonnor-Melvin (BM) \cite{Bonnor:1954tis,Melvin:1963qx} and KBR solutions in the appendix.

\section{Local solution}

The metric of our new rotating black hole in external E/M fields is
\bea
ds^2&=&-\frac{Q }{\mathbf{L} \Omega ^4}\Big[a x^2 d\phi+\frac{P r^2 (1-x^2) \mathbf{L} \Omega ^2-a^2 x^2 \mathbf{H} }{P r^4
   (1-x^2)-a^2  x^4Q} dt\Big]^2\cr
   &&+\frac{ (1-x^2)P}{\mathbf{L} \Omega ^4}\Big[
   r^2 d\phi+\frac{a(Q x^2 \mathbf{L} \Omega ^2-r^2 \mathbf{H}) }{P r^4 (1-x^2)-a^2  x^4Q}dt\Big]^2 +\mathbf{L}\Big(\frac{dr^2}{Q}+\frac{dx^2}{(1-x^2)P}\Big)\,,\nn\\
\mathbf{L}&=&\frac{1}{\Omega ^4 I_1^2}\Big(I_1^2(1+\mathcal{E}^4a^2 r^2 x^2 ) \Sigma  + \mathcal{E}^2 \mu  I_2 r^3 x^2\big[ \mathcal{E}^2 \mu
   I_2r x^2-2I_1 (1+\mathcal{E}^2a^2 x^2 ) \big]\Big)\,,\cr
\mathbf{H}&=& x^2Q+P r^2 (1-x^2)\,,\qquad I_1=1+\ft{1}{2}\mathcal{E}^2a^2\,,\qquad I_2=1+\mathcal{E}^2a^2\,,\cr
\Sigma&=&r^2+a^2 x^2\,,\qquad P=1-\mathcal{E}^2 \Big(\mu ^2\frac{ I_2}{I_1^2}-a^2\Big) x^2\,,\qquad Q=(1-\mathcal{E}^2r^2 ) \Delta\,,\cr
\Omega^2&=&1-\mathcal{E}^2r^2 +\mathcal{E}^2x^2  \Delta\,,\qquad \Delta=\Big(1+\mathcal{E}^2 \mu ^2\frac{ I_2}{I_1^2}\Big) r^2-2 \mu\frac{  I_2 }{I_1}r+a^2\,.
\eea
The Maxwell potential for electric or magnetic sources is respectively given by
\bea
A_{\1}^{\mathrm{ele}}&=&\frac{1}{\mathcal{E}\Omega ^2  \mathbf{L}}\Bigg\{
ax\partial_r\Omega\Big[
   r^2 d\phi+\frac{a(Q x^2 \mathbf{L} \Omega ^2-r^2 \mathbf{H}) }{P r^4 (1-x^2)-a^2  x^4Q} dt\Big]\cr
   &&-r\partial_x\Omega\Big[a x^2 d\phi+\frac{P r^2 (1-x^2) \mathbf{L} \Omega ^2-a^2 x^2 \mathbf{H} }{P r^4 (1-x^2)-a^2  x^4Q}dt\Big]
\Bigg\}\,,\cr
A_{\1}^{\mathrm{mag}}&=&-\frac{a (1-\mathcal{E}^2 r^2)}{ \mathcal{E}\mathbf{L} \Omega ^{3}}\Big[a x^2 d\phi+\frac{P r^2 (1-x^2) \mathbf{L} \Omega ^2-a^2 x^2 \mathbf{H} }{P r^4
   (1-x^2)-a^2  x^4Q}dt\Big]\cr
   &&-\frac{1- \mathcal{E}^2 \big( \mu \frac{ I_2}{I_1}r-a^2\big)x^2}{\mathcal{E}\mathbf{L}  \Omega ^{3}}\Big[
   r^2 d\phi+\frac{a(Q x^2 \mathbf{L} \Omega ^2-r^2 \mathbf{H}) }{P r^4 (1-x^2)-a^2  x^4Q} dt\Big]+\frac{d\phi}{\mathcal{E}}\,.
\label{LH black hole}
\eea
They satisfy $F_{\2}^{\mathrm{ele}}=*F_{\2}^{\mathrm{mag}}$. Mixed E/M fields can be obtained via the electromagnetic duality. Unlike the type-D PO solution \cite{Podolsky:2025tle,Ovcharenko:2025cpm}, ours belongs to Petrov type I. In the limit \(a = 0\), the rotating solution becomes static
\bea
ds^2&=&\frac{(1-\mathcal{E}^2\mu  r x^2 )^2}{\Omega ^4}\Big[-\frac{Q}{r^2}dt^2+\frac{r^2}{Q}dr^2+\frac{r^2}{P}\frac{dx^2}{1-x^2}
\Big]+\frac{P r^2 (1-x^2)}{(1-\mathcal{E}^2\mu  r x^2 )^2}d\phi^2\,,\cr
A_{\1}^{\mathrm{ele}}&=&-\frac{\mathcal{E}Q x }{r \Omega  (1-\mathcal{E}^2r^2 )}dt\,,\qquad \hbox{or}\qquad
A_{\1}^{\mathrm{mag}}=-\frac{1}{\mathcal{E}}\Big(\frac{\Omega}{1-\mathcal{E}^2\mu  r x^2}-1\Big)d\phi\,.\label{L-H BH static}
\eea
Further setting \(\mu = 0\), we obtain a new electrovacuum spacetime background
\bea
ds^2&=&\frac{1}{\big(1- \mathcal{E}^2 r^2
   (1-x^2)\big)^2}\Big[-(1-\mathcal{E}^2 r^2) dt^2+\frac{dr^2}{1-\mathcal{E}^2 r^2}+r^2\frac{ dx^2}{1-x^2}\Big]+r^2 (1-x^2) d\phi^2\,,\cr
A_{\1}^{\mathrm{ele}}&=&-\frac{\mathcal{E}r x }{\sqrt{1-\mathcal{E}^2r^2 (1-x^2) }}dt\,,\qquad\hbox{or}\qquad   A_{\1}^{\mathrm{mag}}= -\frac{\sqrt{1-\mathcal{E}^2 r^2 (1-x^2)}-1}{\mathcal{E}}d\phi\,.\label{vacuum}
\eea
The identity $\mathrm{Riem}^2 =56 \mathcal{E}^4 \big(1- \mathcal{E}^2 r^2 (1-x^2)\big)^4= 14  (F^2)^2$ indicates that our electrovacuum background is distinct from both
the BM and Bertotti-Robinson (BR) \cite{Bertotti:1959pf,Robinson:1959ev} universes, but locally belongs to Kundt class of Petrov type D \cite{Ovcharenko:2026uxi}.

\section{A brief analysis of the global structure}

The metric \eqref{LH black hole} is degenerate at the north and south poles ($x=\pm1$). To eliminate the naked closed timelike curves and conical singularities, we perform the coordinate transformation
\be
\phi\quad\rightarrow\quad\phi'=\frac{\phi}{P_0}\,,\qquad t\quad\rightarrow\quad t'=t-a\frac{\phi}{P_0}\,,\qquad P_0=1+\mathcal{E}^2 \Big(a^2-\mu ^2\frac{I_2 }{I_1^2}\Big)\,,\label{CTC and coinical singularity}
\ee
so that the azimuthal angle $\phi$ has the standard $2\pi$ period. The black hole horizons are determined by the roots of \(Q(r)\). In addition to the inner and outer horizons \(r_\pm\), the solution also admits a further Cauchy horizon \(r_c\)
\be
r_\pm=\frac{\mu  I_2\pm\sqrt{\mu ^2 I_2-a^2 I_1^2} }{I_1^2+\mathcal{E}^2 \mu ^2 I_2}I_1\,,\qquad r_c=\frac{1}{\mathcal{E}}\,.
\ee
We observe that
\be
r_c-r_+=\frac{I_2 (I_1-\mu  \mathcal{E})^2 (I_1 I_2 (I_1+\mu  \mathcal{E})+m \mathcal{E} (I_1+m \mathcal{E}))}{\mathcal{E} (I_1+m \mathcal{E})^2 \left(I_1^2+I_2 \mu ^2 \mathcal{E}^2\right)}\ge 0\,,
\ee
where $m=\sqrt{\mu ^2 I_2-a^2 I_1^2}\geq0$. Thus, the Cauchy horizon \(r_c\) play a role analogous to the cosmological horizon in dS spacetime, but with a key difference: the event horizon can never swallow the cosmological horizon in our solution.

We now analyse the global structure, focusing on the simpler static case, namely the Schwarzschild black hole immersed in the new electrovacuum. The electrovacuum \eqref{vacuum} is globally distinct from the Kundt class since it is uncharged, whilst the latter are not electrovacua but spacetimes generated by (accelerating) electric and magnetic charges \cite{Ovcharenko:2026uxi}. See \cite{Ma:2026uok} for their global differences. For $\mu>0$, a black hole emerges. The temperature and entropy associated with the event and cosmological horizons are given by
\bea
S_+&=&\frac{A_+}{4}=\frac{4 \pi  \mu ^2}{P_0(1-\mathcal{E}^2 \mu ^2)^2}\,,\qquad\qquad\quad\quad\,\,
T_+=\frac{(1-\mathcal{E}^2 \mu ^2)^2}{8 \pi  \mu }\,,\nn\\
S_c &=& \frac{A_c}{4}=\frac{\pi }{2 P_0\mathcal{E}^2 (1-\mathcal{E} \mu )^2}\int_{-1}^1\frac{1}{x^2}dx\,,\qquad T_c=\frac{\mathcal{E} (1-\mathcal{E} \mu )^2}{2 \pi }\,.
\eea
Unlike dS spacetime, the area of the cosmological horizon is non-compact. A clear picture emerges. For $\mu=0$, the electrovacuum background \eqref{vacuum} describes a universe surrounded by the cosmological horizon, beyond which the metric becomes cosmological, with a spacelike curvature singularity at $r\rightarrow \infty$. When $\mu$ is turned on, a black hole with event horizon $r_+$ forms inside the cosmological horizon. The spacetime is static in $r_+<r<r_c$. The Riemann curvature invariants indicate that singularities are located at both $r=0$ and \(W\equiv 1 - \mathcal{E}^2 \mu r x^2 = 0\), while the $r=\infty$ region becomes regular for non-vanishing ${\cal E}\mu$. The $r=0$ singularity is hidden inside the event horizon, whilst the {\it cosmological singularity} $W=0$ is beyond the cosmological horizon provided that ${\cal E} \mu<1$. For ${\cal E} \mu\ge 1$, the $W=0$ singularity can appear inside the cosmological horizon and becomes naked. In particular, when ${\cal E} \mu= 1$, the event horizon and cosmological horizon merge to become a singular hypersurface, in which case the entropy $S_+$ diverges.

\begin{figure}[hbtp]
  \centering
  \includegraphics[width=0.4\textwidth]{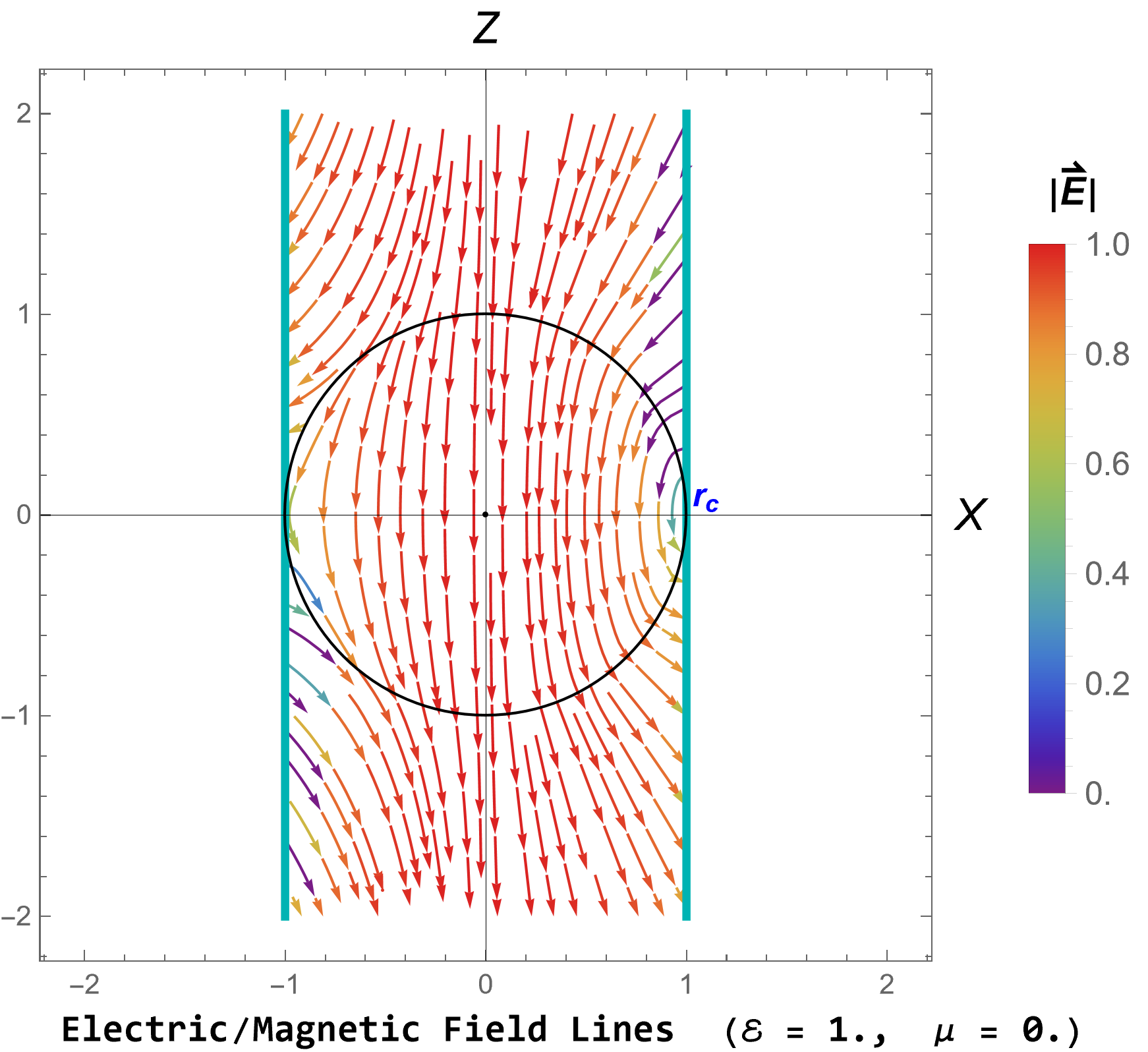}\ \
  \includegraphics[width=0.4\textwidth]{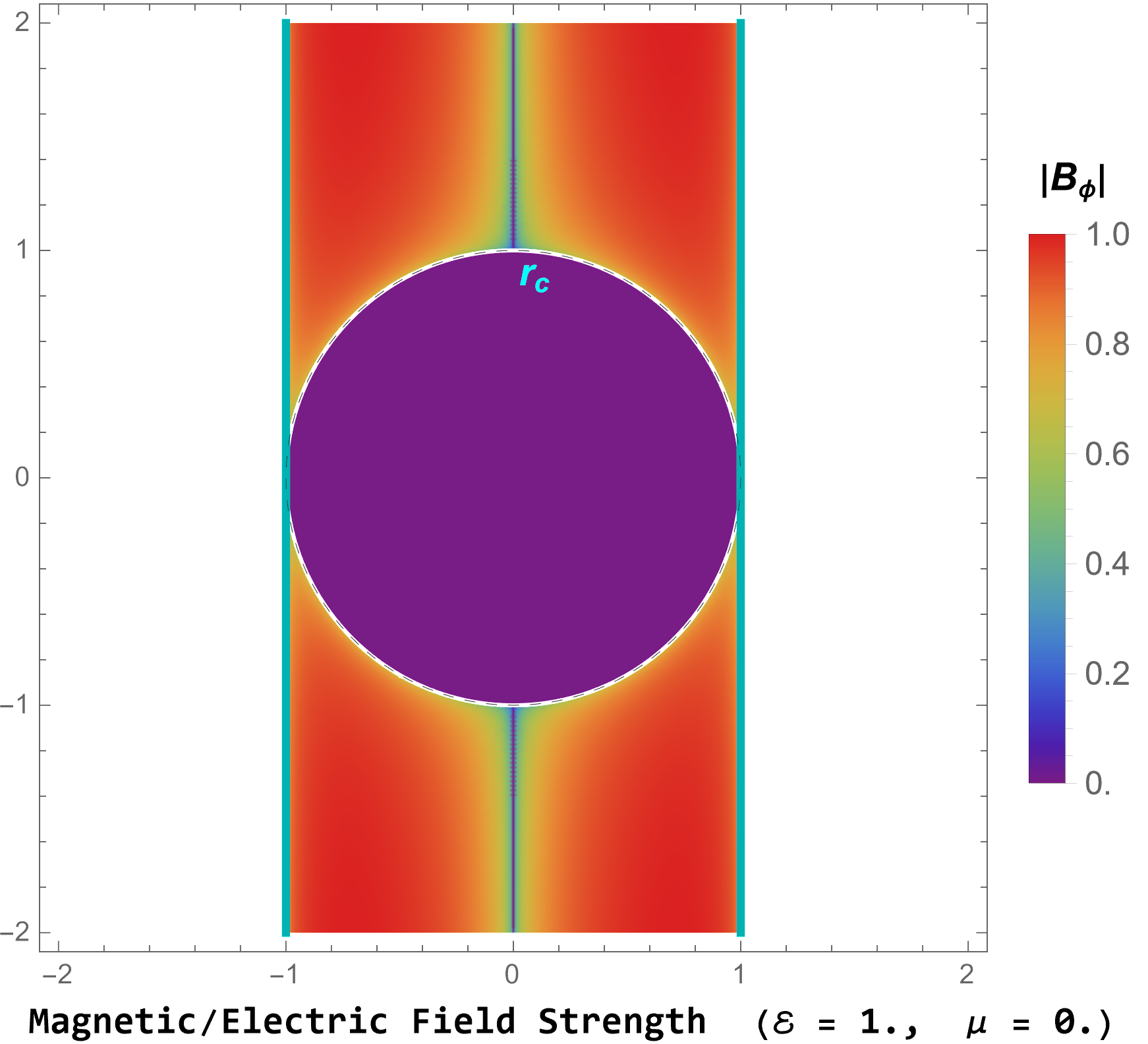}\
  \includegraphics[width=0.4\textwidth]{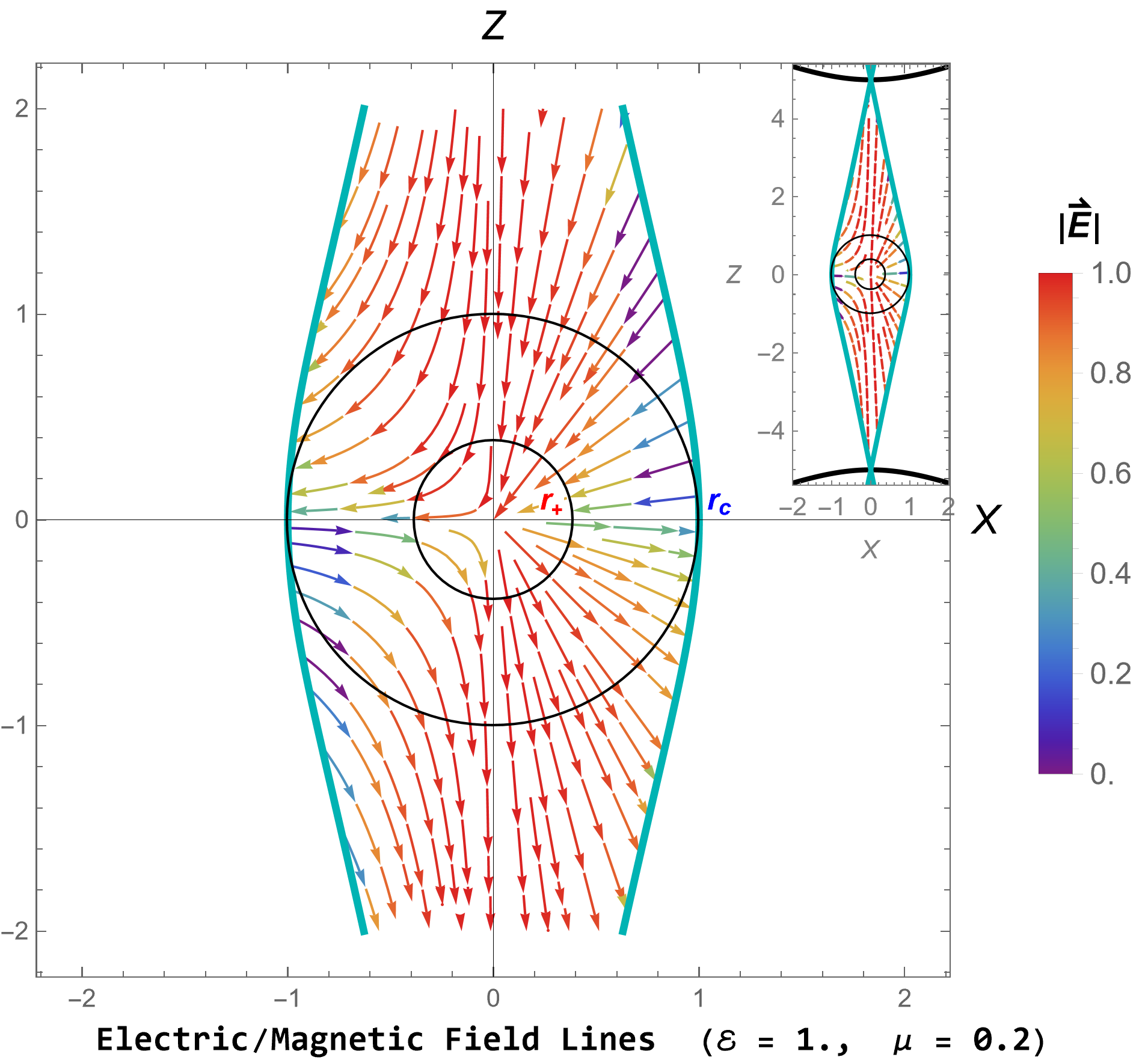}\ \
  \includegraphics[width=0.4\textwidth]{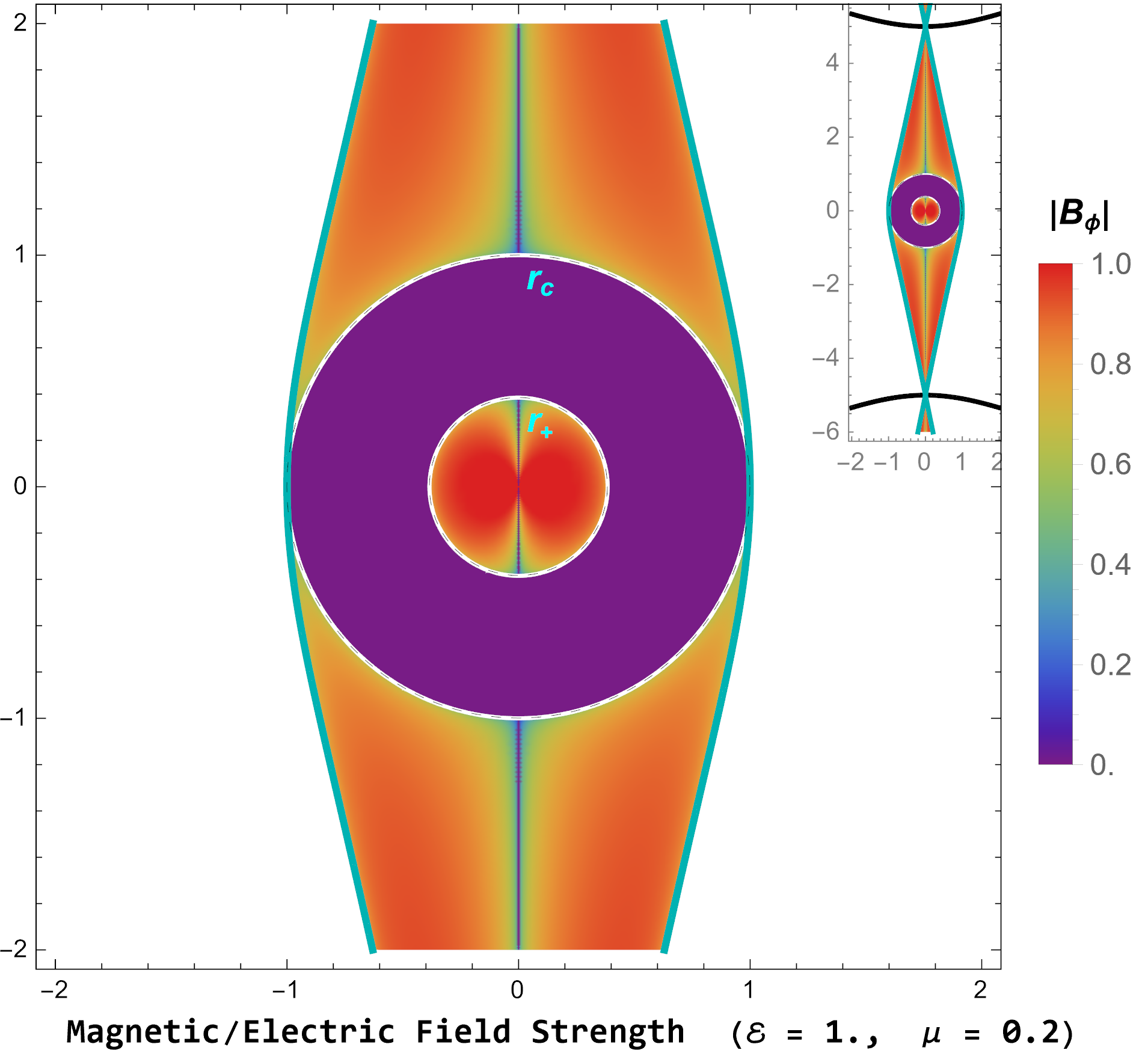}\
  \includegraphics[width=0.4\textwidth]{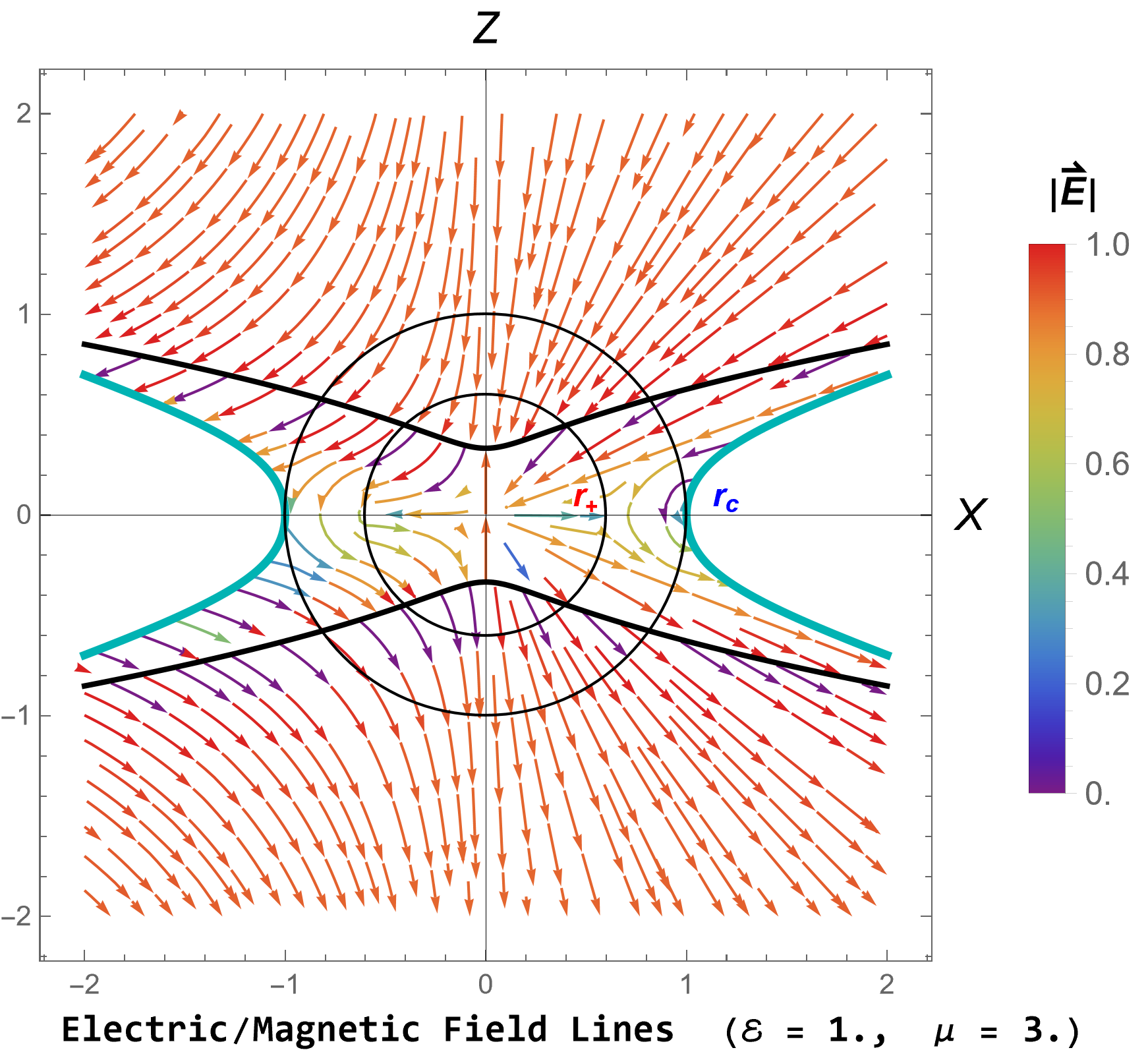}\
  \includegraphics[width=0.4\textwidth]{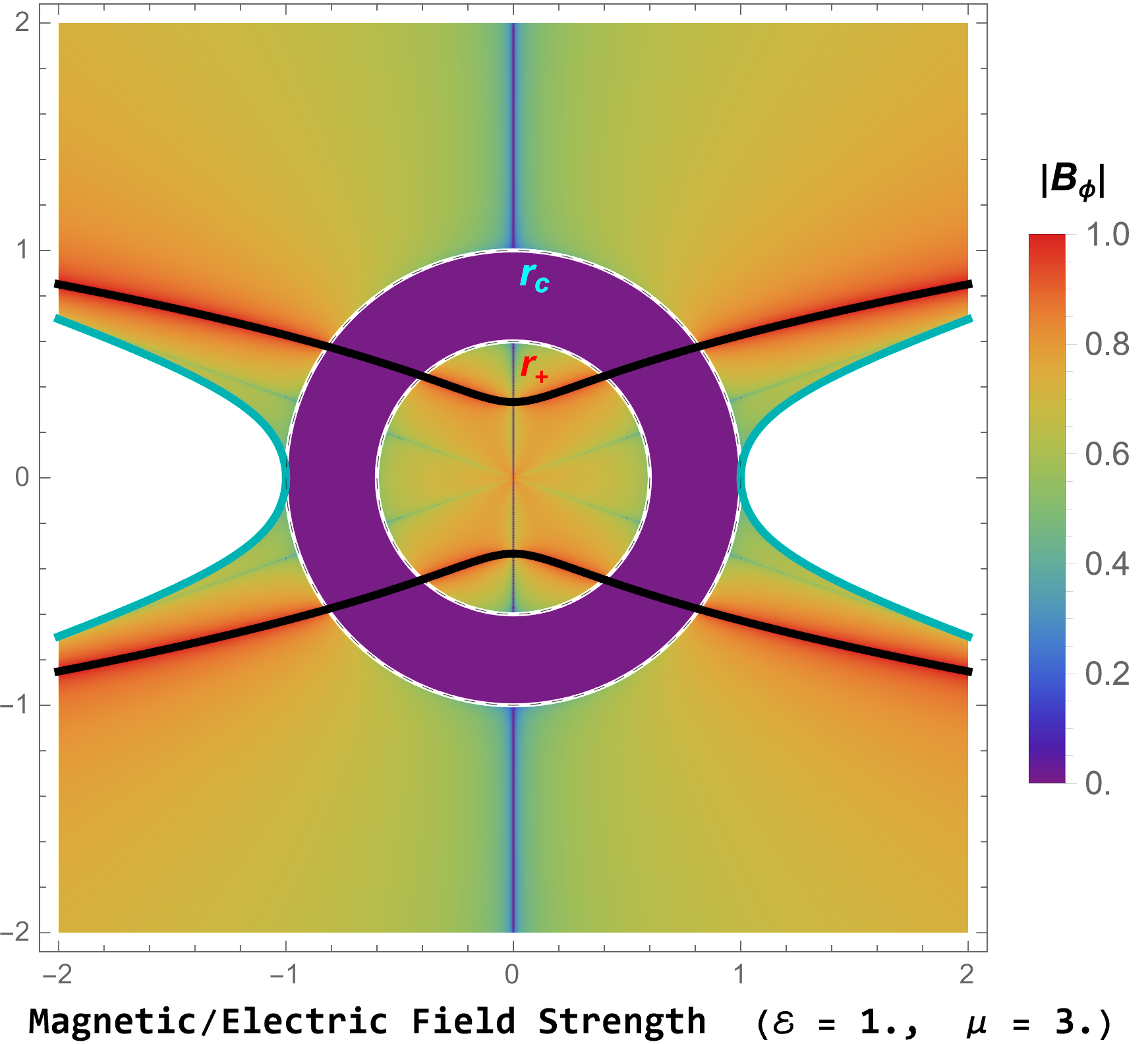}\
 \caption{\small We present the E/M field lines, together with the corresponding spacetime structure for the vacuum case, as well as for the two regimes \(\mathcal{E}\mu < 1\) and \(\mathcal{E}\mu > 1\). We adopt the rescaling prescription \(\{|\vec E|,|\vec B|\} \rightarrow (\frac{\log(1+p\{|\vec{E}|,B_\phi\})}{\log(1+p\{|\vec{E}|,B_\phi\}_{
 \mathrm{max}})})^{0.4}\), which maps the E/M field strengths into the interval \([0,1]\) while preserving their ordering. This enables a visualization of the field magnitude using a rainbow colormap. In particular, we set \(p_E = 8\) for the electric field and \(p_B = 50\) for the magnetic field. Both the event and cosmological horizons are apparent. The black lines represent the cosmological singularities, while the white region outside the cosmological horizon is where the Maxwell field becomes purely imaginary and hence phantomlike.}\label{LH electric}
\end{figure}

The understanding of the spacetime is incomplete without analysing the E/M fields. One important feature is that there exists negative $\Omega^2$ beyond the cosmological horizon, in which region the Maxwell potential $A_\1$ becomes purely imaginary and hence phantomlike. Our solution indicates that there is no geometric barrier between normal and phantomlike fields and they can continuously transform into each other across a smooth spacetime boundary. In Fig.~\ref{LH electric}, we present both the E/M field lines and the
spacetime structure. The procedure of computing the E/M field components is presented in Appendix \ref{electric and magnetic field}. In all the graphs, we set ${\cal E}=1$. We present three cases, $\mu=0$, $\mu<1$ and $\mu>1$. We consider the case where the static region involves only the electric fields, in  which case, a magnetic component emerges in the azimuthal direction outside the cosmological horizon where the radial $r$ becomes timelike. The black solid line represents the cosmological singularity at $W=0$, which lies outside the cosmological horizon for $\mu<1$, but enters it for $\mu>1$. The white region outside the cosmological horizon is where the Maxwell field is purely imaginary and hence phantomlike. The cyan solid line separates the normal Maxwell field from the phantomlike field.

\section{Black hole thermodynamics}

The black hole metric \eqref{LH black hole} is not asymptotically Minkowski; therefore, we do not have a canonical scaling for the time coordinate. Working in the coordinate system \eqref{CTC and coinical singularity}, we follow the approach of \cite{Ma:2026otg} and make an additional coordinate transformation and a $U(1)$ gauge transformation
\bea
t\rightarrow t'=\lambda_1 t\,,\quad \phi\rightarrow\phi'=\phi+\mathcal{E}\lambda_2t\,;\qquad A_{\1}^{\text{mag}}\rightarrow A_{\1}^{\text{mag}}+\lambda_3dt\,,\quad A_{\1}^{\text{ele}}\rightarrow A_{\1}^{\text{ele}}+\lambda_4dt\,.\label{coord thermodynamics}
\eea
For definiteness, we focus on the case of an external electric field in the following discussion. The corresponding magnetic solution follows directly from the electromagnetic duality.

In Einstein gravity, the entropy of the black hole \eqref{LH black hole} is simply one quarter of the event horizon area. Owing to rotation, the external electric field generates a non-vanishing magnetic charge \(Q_m\). We therefore have
\bea
S=\frac{\pi  (r_+^2+a^2)}{P_0 (1-\mathcal{E}^2 r_+^2)}\,,\qquad Q_m=\frac{1}{4\pi}\oint_{S^2}F_{\2}^{\mathrm{ele}}=\frac{a \mathcal{E} \mu  \sqrt{I_2}}{I_1 P_0}\,.
\eea
The spacetime admits two commuting Killing vector fields of the form \(\xi = c_1 \partial_t - c_2 \partial_\phi\), which generate the black hole mass \(M\) and angular momentum \(J\), respectively. Since the black hole \eqref{LH black hole} carries magnetic charge, we adopt the Komar 2-form  that is explicitly invariant under electromagnetic duality \cite{Liu:2022wku}
\be
c_1M+2c_2J=\frac{1}{8\pi}\oint_{S^2}\mathbf{Q}\,,\qquad
\mathbf{Q}=-{*d}\xi-2{*F}_{\2}^{\mathrm{ele}}(i_\xi A_{\1}^{\mathrm{ele}})-2{*F}_{\2}^{\mathrm{mag}}(i_\xi A_{\1}^{\mathrm{mag}})\,.
\ee
The Komar integration yields the black hole mass and angular momentum
\be
M=\lambda _1\frac{ \mu  I_2}{I_1 P_0}-2 \mathcal{E}\lambda _2 J-\lambda _3 Q_m\,,\qquad J=\frac{ \mu  a\sqrt{I_2} (P_0+\sqrt{I_2})}{2 I_1 P_0^2}\,,
\ee
It is evident that for the four conserved charges, \(S\), \(J\) and \(Q_m\) are independent of the parameters \(\lambda_{i=1\sim4}\), whereas the mass is. The corresponding thermodynamic potentials are given by temperature, angular velocity and magnetic potential, each of which depends on the $\lambda_i$:
\bea
T &=& \frac{(1-\mathcal{E}^2 r_+^2) (I_2 r_+^2-a^2) \lambda _1}{2 \pi  r_+ (r_+^2+a^2)
   \big(1+\sqrt{I_2(1-\mathcal{E}^2 r_+^2)}\big)}\,,\cr
\omega &=&\frac{a P_0 \lambda _1}{r_+^2+a^2}-\mathcal{E}\lambda _2\,,\qquad
\Phi_m=\frac{a (\sqrt{1-\mathcal{E}^2 r_+^2}-P_0) }{\mathcal{E} (r_+^2+a^2)}\lambda _1-\lambda _3\,.
\eea
These thermodynamic quantities satisfy the Smarr relation $M=2TS+2\omega J+\Phi_mQ_m$, regardless of the values of $\lambda_i$. To determine the parameters $\lambda_i$ from the first law $dM=Td S+\omega d J+\Phi_m d Q_m$, we introduce two parameters $(\alpha,\beta)$, defined by $a=\sqrt{\alpha^2-1}/\mathcal{E}$ and $ r_+=\sqrt{1-\beta^2}/\mathcal{E}$ so that $\lambda_i=\lambda_i(\alpha,\beta,{\cal E})$. We find
\bea
\lambda_1&=&\frac{\sqrt{\alpha ^2-1} \beta  (\alpha ^2-\beta ^2) \big((1+\alpha ^2) \beta -2 \alpha \big)}{\alpha  (\alpha  \beta
   -1) (1-\beta ^2) \gamma }\Big[
   (1-\beta ^2) \big(\alpha ^3-(1-2 \alpha ^3) \beta -(1+3 \alpha ^2+\alpha ^3) \beta ^2\cr
   &&-(\alpha ^3-3
   \alpha -1) \beta ^3\big)c_m^{(0,1)}
   -(\alpha ^2-1) (\alpha  \beta -1) \big(\alpha  (1+2 \alpha )-2 \alpha  \beta +(1-\alpha -\alpha ^2) \beta ^2\big)c_m^{(1,0)}
   \Big]\,,\cr
\lambda_2&=&\frac{\beta  \big((1+\alpha ^2) \beta -2 \alpha \big)}{\alpha  (\alpha  \beta -1) (1-\beta ^2)^2 \gamma }\Bigg\{
\beta  (1-\beta ^2) \Big[2 \alpha ^2 (1+\alpha ^4)-\alpha  (2+3 \alpha ^2+4 \alpha ^3+2 \alpha ^4-4 \alpha
   ^5+\alpha ^6) \beta \cr
   &&-(1-2 \alpha -6 \alpha ^3+\alpha ^4+6 \alpha ^5+2 \alpha ^6+2 \alpha ^7) \beta ^2+(-1+6 \alpha -9
   \alpha ^2+5 \alpha ^3+9 \alpha ^4+4 \alpha ^5+\alpha ^6\cr
   &&+\alpha ^7) \beta ^3+(1+3 \alpha -6 \alpha ^2-\alpha ^3+\alpha ^4-3
   \alpha ^5+\alpha ^7) \beta ^4-2 \alpha  (2+\alpha ^2+\alpha ^4) \beta ^5+4 \alpha ^2 \beta ^6\Big]c_m^{(0,1)}\cr
   &&
   +(\alpha ^2-1) (\alpha  \beta -1)\Big[
   2 \alpha ^3-2 \alpha ^2 (1-2 \alpha +\alpha ^2+2 \alpha ^3) \beta +\alpha  (-1-6 \alpha-2 \alpha ^2+4 \alpha ^3+\alpha ^4 \cr
   &&+2
   \alpha ^5) \beta ^2+2 \alpha  (2+3 \alpha -2 \alpha ^2+\alpha ^3) \beta ^3+(-1+\alpha +\alpha ^2-2 \alpha ^3+\alpha
   ^4-\alpha ^5-\alpha ^6) \beta ^4-4 \alpha ^2 \beta ^5\cr
   &&+2 \alpha ^3 \beta ^6\Big]c_m^{(1,0)}
\Bigg\}\,,\qquad
\lambda_3= c_m+\frac{\alpha  }{\sqrt{\alpha ^2-1}}\lambda _1+\frac{\alpha +2 \alpha  \beta -(1+\alpha +\alpha ^2) \beta ^2}{\beta  \big((1+\alpha ^2) \beta -2 \alpha \big)}\lambda_2\,,\\
\gamma&=&\alpha ^4+2 \alpha ^2 (1+2 \alpha ^2) \beta -\alpha  (1+\alpha ^2) (5+\alpha +2 \alpha ^2) \beta ^2 +(1+4 \alpha ^2) \beta ^3+\alpha  (\alpha ^4-\alpha ^2+\alpha -2) \beta ^4.\nn
\eea
Here \(c_m = c_m(\alpha,\beta)\) remains undetermined by the first law. It is uniquely fixed by further imposing the relation \cite{Christodoulou:1971pcn}
\be
M^2=\frac{S}{4\pi}+\frac{Q_m^2}{2}+\frac{\pi(Q_m^4+4J^2)}{4S}\,.
\ee
The static case is significantly simpler. We have (Note that $Q_m=0$ now.)
\be
M=\lambda _1\frac{ \mu }{P_0}\,,\qquad T=\lambda _1\frac{(1-\mathcal{E}^2 \mu ^2)^2 }{8 \pi  \mu }\,,\qquad S=\frac{4 \pi  \mu ^2}{P_0(1-\mathcal{E}^2 \mu ^2)^2 }\,.
\ee
The resulting expressions satisfy the Smarr relation \(M = 2TS\). However, to ensure the validity of the first law, \(d M = T\,d S\), we must set \(\lambda_1 = c\,P_0^{-1/2}\), where \(c\) is an arbitrary constant. We find that $c=1$ if we impose the relation \(M^2 = S/(4\pi)\).

\section{Conclusion}

There are two well-known electrovacua and related spacetimes in Einstein-Maxwell gravity: the BM and BR universes, both constructed more than half a century ago. In this paper, we presented a new electrovacuum and the corresponding spacetime background \eqref{vacuum}, which is characterized by a cosmological horizon, whose location is determined by the E/M parameter ${\cal E}\sim \sqrt{|F^2|_{\rm max}}$, for the maximum $|F^2|$ inside the cosmological horizon. The electrovacuum locally belongs to the Kundt class, but it is globally distinct in two aspects: Our \eqref{vacuum} is uncharged and supported by external electromagnetic fields; it admits smooth flat limit. The corresponding Kundt solution is not an electrovacuum but charged spacetime; it either suffers from naked singularities or does not admits a smooth flat limit \cite{Ma:2026uok}. The distinction is analogous to the uncharged BR universe, which can be locally transformed into the charged AdS$_2\times S^2$. We constructed an exact solution describing the Kerr black hole immersed in this background and analysed its global structure and black hole thermodynamics. We illustrated how the E/M field is distributed in the spacetime containing the event horizon and cosmological horizon, as well as the black hole and cosmological singularities. An intriguing feature is that beyond the cosmological horizon, where the solution becomes cosmological, the Maxwell field can continuously evolve across a smooth boundary and become phantomlike, indicating that the distinction between normal and phantom matters can be blurred when coupled to gravity.

The emergence of the cosmological horizon from such orindary matter like the Maxwell fields, without invoking a positive cosmological constant or dark energy, is tantalizing. Observational data indicate that our expanding universe asymptotically approaches a dS spacetime in the late-time universe, corresponding to a cosmological constant of order \(\Lambda \sim 10^{-52}\,\mathrm{m}^{-2}\). If this effect is described using the  electrovacuum configuration \eqref{vacuum} rather than introducing a cosmological constant or dark energy, it is sufficient to introduce a very small electric field parameter \(\mathcal{E} \sim \sqrt{\Lambda} \sim 10^{-26}\,\mathrm{m}^{-1} \approx 10^{-42}\,\mathrm{GeV}\) to reproduce the same asymptotic behavior. Our new solution may therefore indicate a new approach to analysing observational data related to black holes and late-time cosmology in our Universe.

\section*{Acknowledgement}

We are grateful to Run-Qiu Yang for useful discussions. L.M.~is supported in part by National Natural Science Foundation of China (NSFC) grant No.~12447138, Postdoctoral Fellowship Program of CPSF Grant No.~GZC20241211, the China Postdoctoral Science Foundation under Grant No.~2024M762338 and the National Key Research and Development Program No.~2022YFE0134300. H.L.~is supported in part by the NSFC grants No.~12375052 and No.~11935009. Both are also supported in part by the Tianjin University Self-Innovation Fund Extreme Basic Research Project Grant No.~2025XJ21-0007.

\appendix

\section{E/M field in black holes with/out cosmological horizon}\label{electric and magnetic field}

\subsection{With cosmological horizon}

Here we explain how Fig.~\ref{LH electric} is constructed. To describe the E/M field of the black hole \eqref{L-H BH static}, we introduce an orthonormal frame \(ds^2 = \eta_{\underline{a}\underline{b}}\, e^{\underline{a}} e^{\underline{b}}\) and express the 2-form field strength as \(F_{\underline{a}\underline{b}} = F_{\mu\nu}\, E^\mu_{\underline{a}} E^\nu_{\underline{b}}\). For the static metric, the vielbein choice is apparent. Due to the presence of the cosmological horizon, we analyse the regions \(r_+ < r < r_c\) and \(r > r_c\), or black hole interior $r<r_+$ separately. For simplicity, we focus on the external electric component \(A_{\1}^{\rm ele}\); the magnetic dual follows directly from the electromagnetic duality $F_{\2}^{\mathrm{ele}}=*F_{\2}^{\mathrm{mag}}$. In the static region, there is only the electric field:
\bea
r_+ < r < r_c:\quad E_{\underline{1}}&=&F_{\underline{1}\underline{0}}=-\mathcal{E}\frac{x  \Omega ^2}{1-\mathcal{E}^2 \mu r x^2  }\Big[
\frac{2\mu}{1-\mathcal{E}^2 \mu r x^2}\partial_r\Omega+\frac{1+\mathcal{E}^2\mu^2}{\Omega}
\Big]\,,\cr
E_{\underline{2}}&=&F_{\underline{2}\underline{0}}=\mathcal{E}\frac{\sqrt{1-x^2} \sqrt{P Q}  \,\Omega }{r (1-\mathcal{E}^2 \mu  r x^2)^2}\,,\qquad E_{\underline{3}}=F_{\underline{3}\underline{0}}=0\,.
\eea
In the cosmological region, in addition to the electric field, the magnetic fields also appears
\bea
r > r_c\quad \hbox{or}\quad r<r_+:\qquad E_{\underline{1}}&=&F_{\underline{1}\underline{0}}=\mathcal{E}\frac{x  \Omega ^2}{1-\mathcal{E}^2 \mu r x^2  }\Big[
\frac{2\mu}{1-\mathcal{E}^2 \mu r x^2}\partial_r\Omega+\frac{1+\mathcal{E}^2\mu^2}{\Omega}
\Big]\,,\cr
   B_{\underline{3}}&=&F_{\underline{1}\underline{2}}=\mathcal{E}\frac{\sqrt{1-x^2} \sqrt{ -PQ}\,  \Omega }{r (1-\mathcal{E}^2 \mu  r x^2)^2}\,.
\eea
Consider a cross section at a fixed azimuthal angle \(\phi = 0, 2\pi\), the spherical coordinates \(\{r, x\}\) are related to the Cartesian coordinates \(\{X, Z\}\) via \(X = r \sqrt{1 - x^2}\) and \(Z = r x\). Accordingly, the components of the electric field \(E_{\underline{1},\underline{2}}\) in the orthonormal frame can be expressed in the Cartesian coordinates as follows
\bea
E_X=\sqrt{1-x^2}E_{\underline{1}}+xE_{\underline{2}}\,,\qquad E_Z=xE_{\underline{1}}-\sqrt{1-x^2}E_{\underline{2}}\,.\label{electric field in XZ}
\eea
The magnetic field is always orthogonal to the \(X-Z\) plane; therefore, we only plot the magnetic field strength \(B_\phi=B_{\underline{3}}\) in the \(X-Z\) plane.

In addition, we pay attention to two important curves. The first \( W=1-\mathcal{E}^2\mu r x^2=0\), where the curvature singularities resides. The second is \(\Omega^2=0\), which specifies the boundary beyond which the Maxwell field becomes purely imaginary and hence phantomlike. In the \(X-Z\) coordinates, these two curves are given as follows
\bea
\mathcal{E}^2 \mu\frac{  Z^2}{\sqrt{X^2+Z^2}}-1=0\,,\qquad 1-\mathcal{E}^2 X^2+\mathcal{E}^2 \mu  Z^2 \Big(\mathcal{E}^2 \mu -\frac{2}{\sqrt{X^2+Z^2}}\Big)=0\,.
\eea

\subsection{Without cosmological horizon}

\begin{figure}[hbtp]
  \centering
  \includegraphics[width=0.4\textwidth]{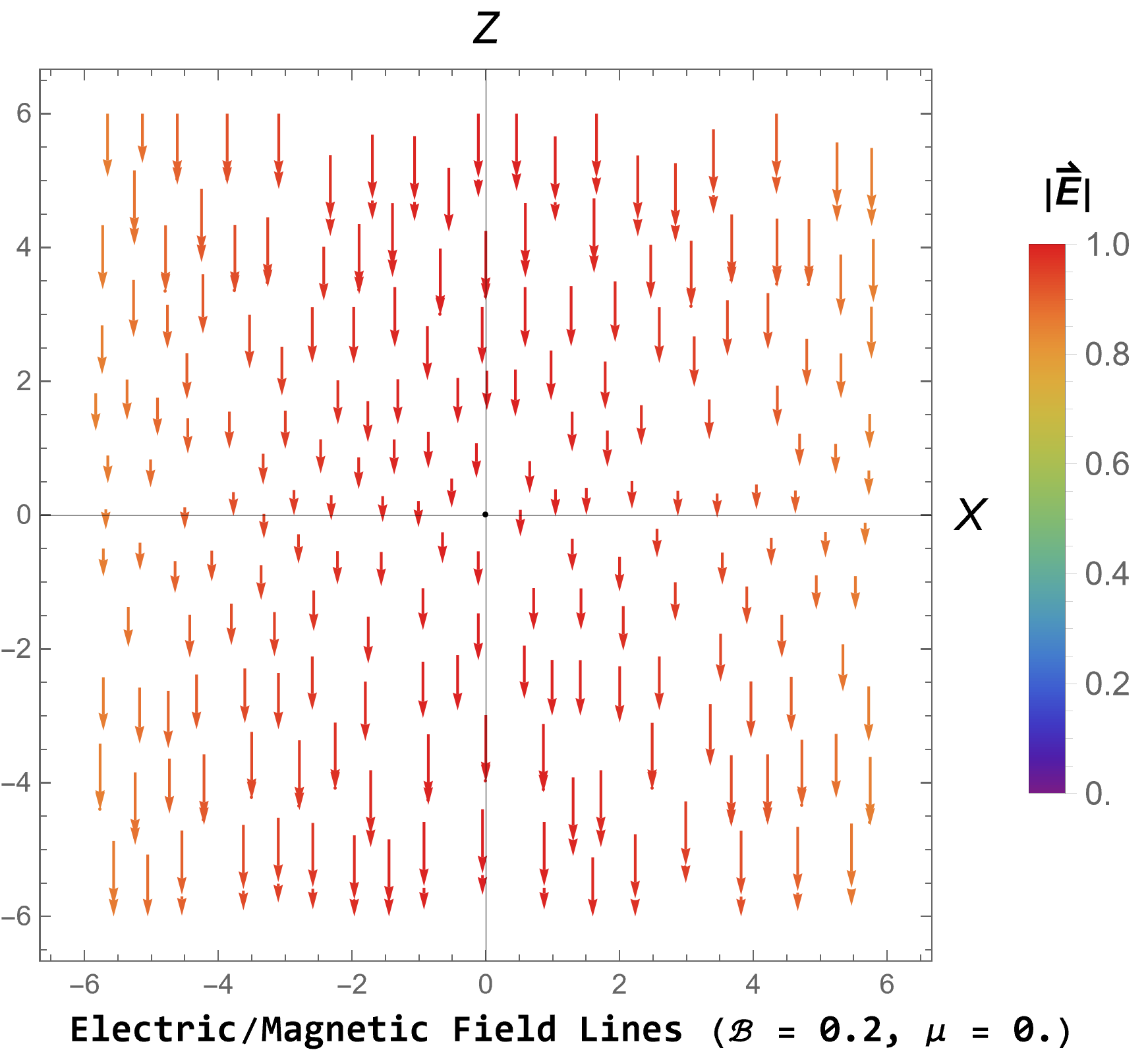}\
  \includegraphics[width=0.4\textwidth]{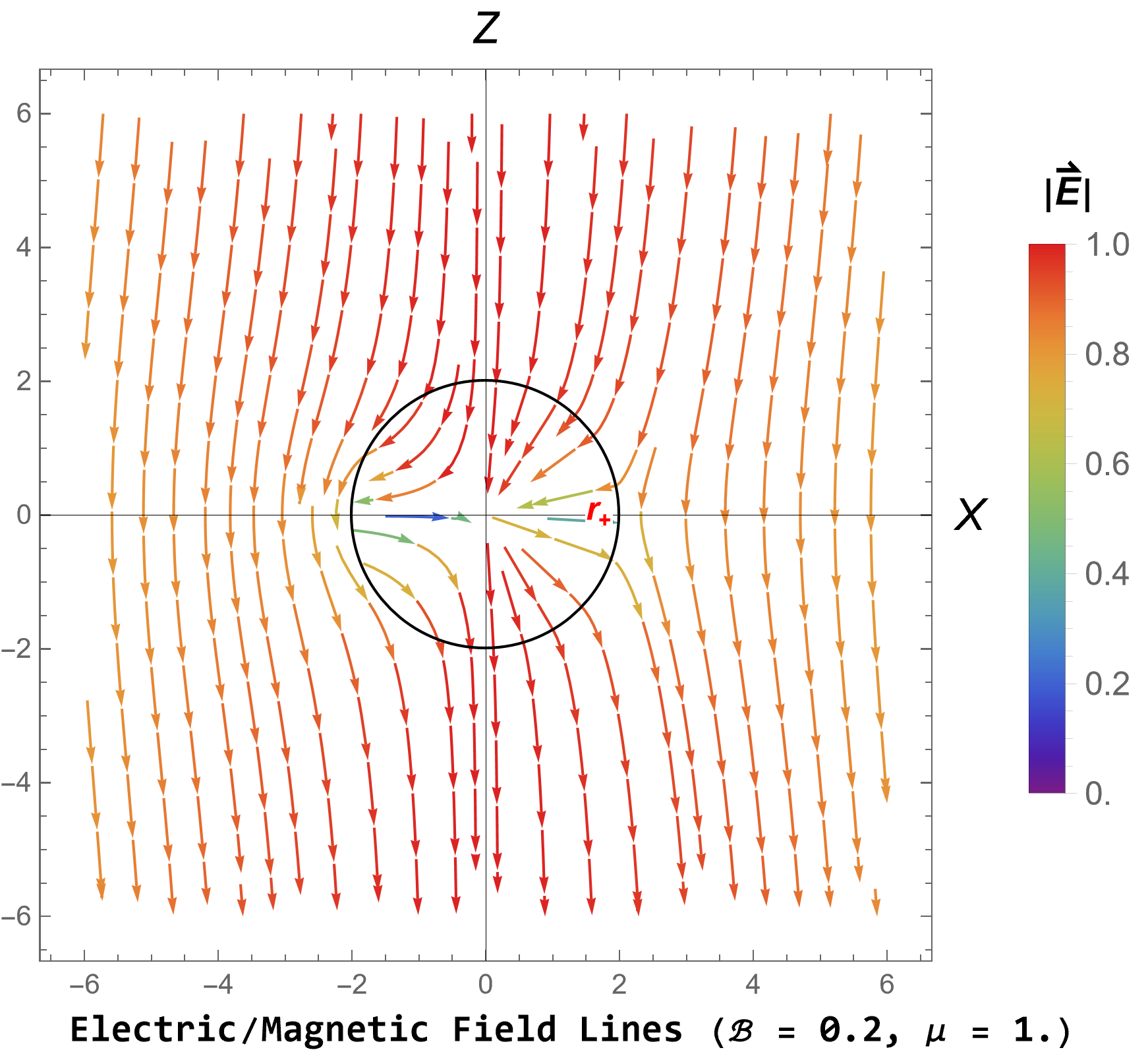}\
  \includegraphics[width=0.4\textwidth]{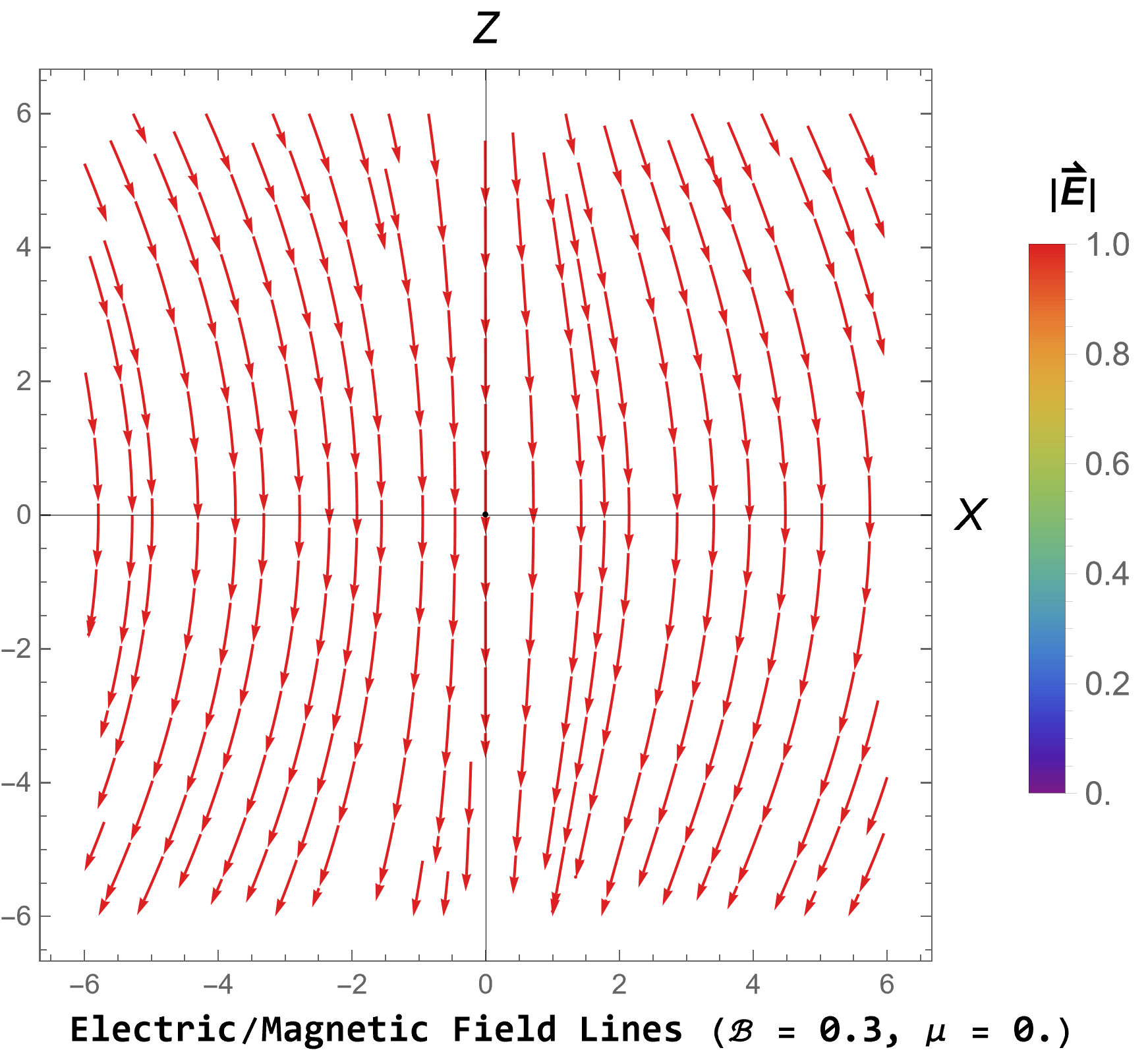}\
  \includegraphics[width=0.4\textwidth]{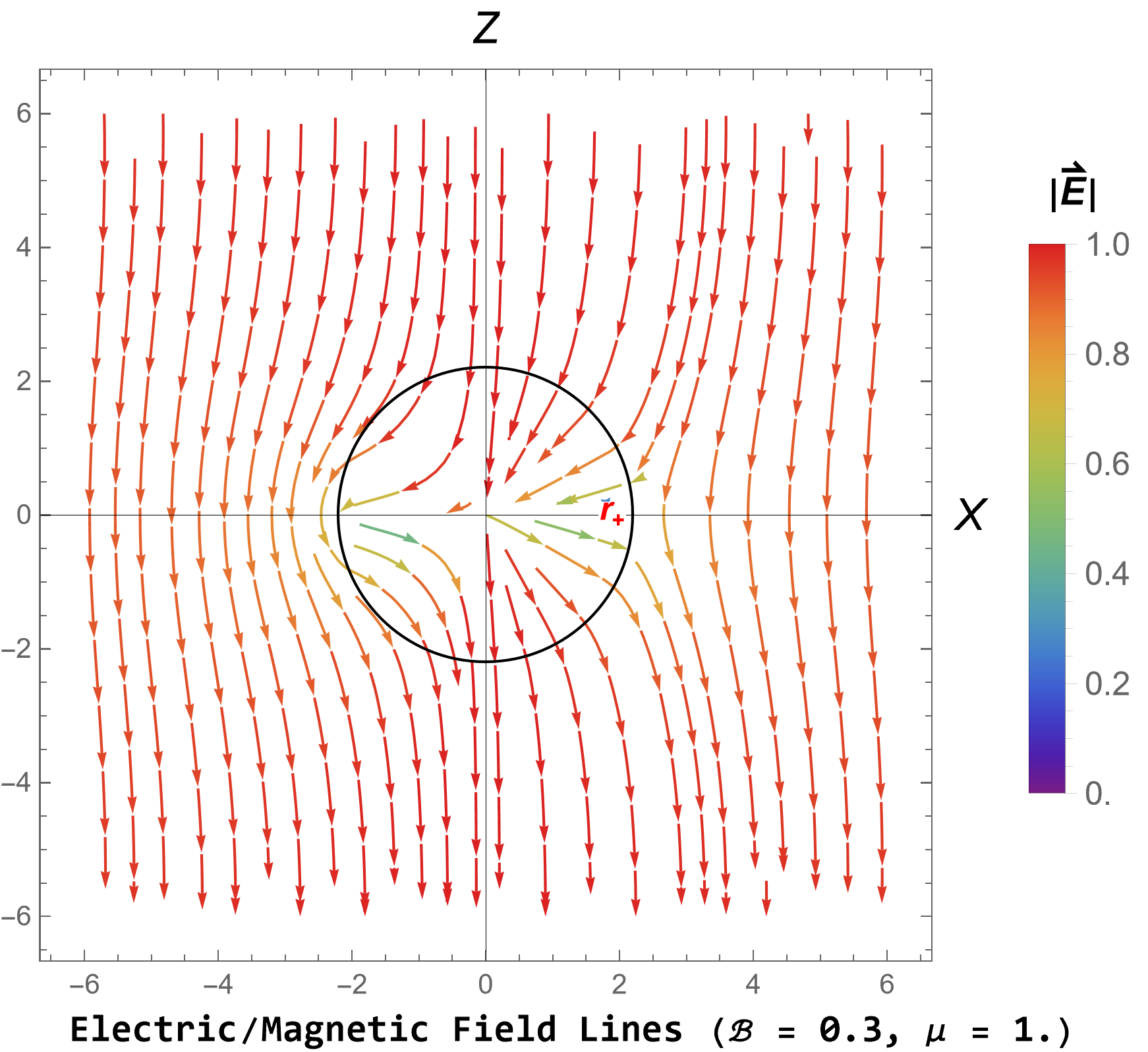}\
  \includegraphics[width=0.4\textwidth]{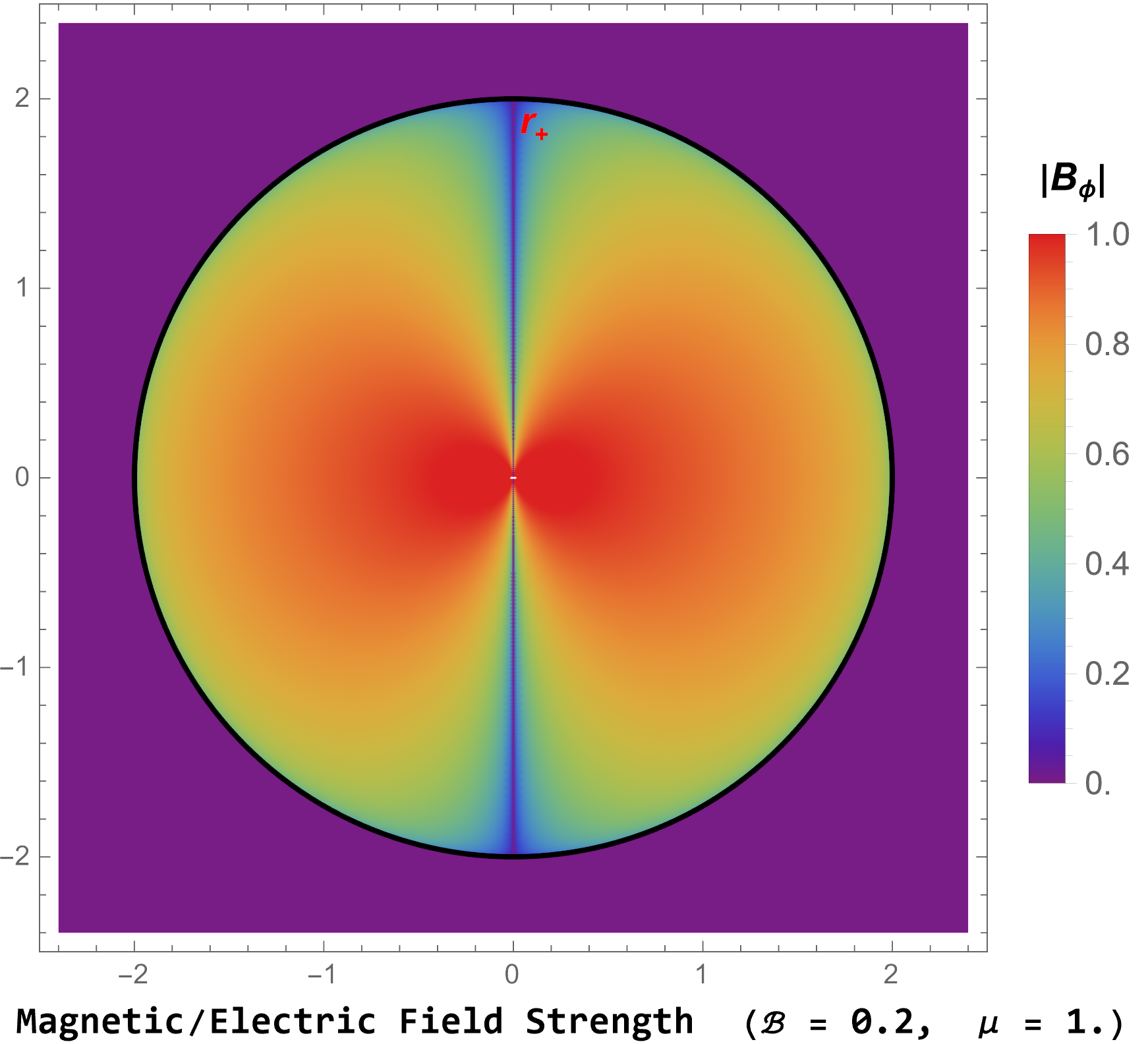}\
  \includegraphics[width=0.4\textwidth]{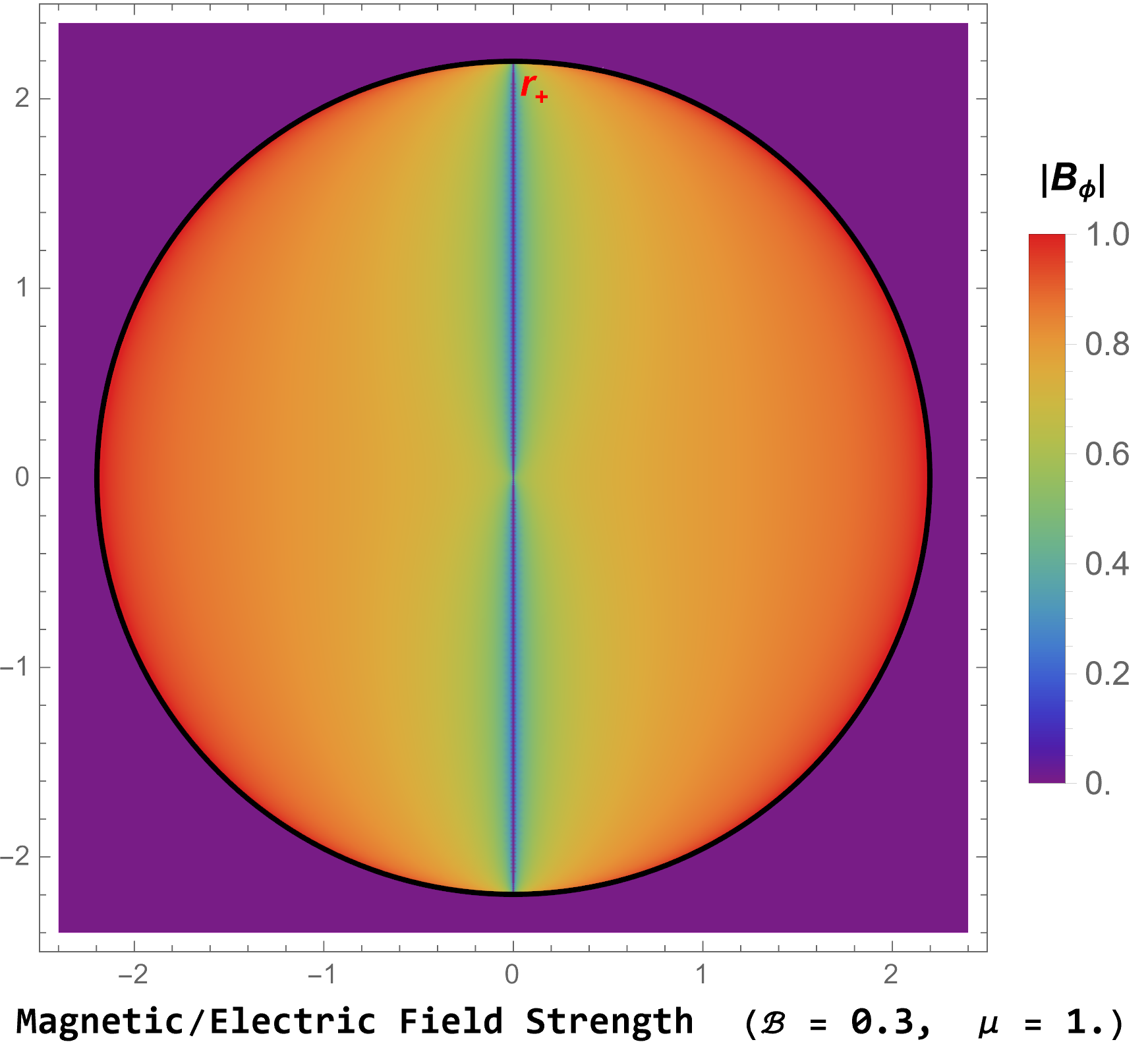}\
 \caption{\small The top two panels display the field lines of the SM solution, while the middle two panels show the field lines of the static PO black hole. The left top two panels correspond to the vacuum, whereas the right top two panels show black holes with an event horizon. The bottom two panels illustrate the magnetic field strength in the interior region of SM (left) and PO (right) black holes.}\label{Sch electric}
\end{figure}

For comparison, we consider the Schwarzschild-Melvin (SM) and static PO black holes. The solutions are
\bea
\hbox{SM}:&& ds^2=\Big(1+\frac{\mathcal{B}^2}{4}  r^2 (1-x^2)\Big)^2 \Big(-f dt^2+\frac{dr^2}{f}+\frac{r^2
   dx^2}{1-x^2}\Big)+\frac{r^2 (1-x^2) d\phi^2}{\Big(1+\frac{\mathcal{B}^2}{4} r^2
   (1-x^2)\Big)^2}\,,\cr
&&A_{\1}^{\mathrm{ele}}=-\mathcal{B} f r xdt\,,\qquad\mathrm{or}\qquad \qquad A_{\1}^{\mathrm{mag}}=\frac{2 \mathcal{B}}{\mathcal{B}^2+\frac{4}{r^2 (1-x^2)}}d\phi\,,\qquad f=1-\frac{2\mu}{r}\,;\label{Sch Melvin}\\
\hbox{PO}: && ds^2=\frac{1}{\Omega^2}\Big[
-\frac{Q}{r^2}dt^2+\frac{r^2}{Q}dr^2+\frac{r^2}{P}\frac{dx^2}{1-x^2}+Pr^2(1-x^2)\frac{d\phi^2}{P_0^2}
\Big]\,,\cr
&&A_{\1}^{\mathrm{ele}}=\frac{\partial_x\Omega}{\mathcal{B}r} dt\,,\qquad \mathrm{or}\qquad A_{\1}^{\mathrm{mag}}=-\frac{r\partial_r\Omega-\Omega+1}{\mathcal{B}}\frac{d\phi}{P_0}\,, \label{PO solution}
\eea
We shall consider the $A_\1^{\rm ele}$ solution for both cases, and the electric fields in vielbein base are given by
\bea
\hbox{SM}:\quad r>r_+\quad && E_{\underline{1}}=-\frac{\mathcal{B} x}{\Big(1+\frac{\mathcal{B}^2}{4}  r^2 (1-x^2)\Big)^2}\,,\quad
E_{\underline{2}}=\frac{\mathcal{B} \sqrt{f} \sqrt{1-x^2}}{\Big(1+\frac{\mathcal{B}^2}{4}  r^2 (1-x^2)\Big)^2}\,,\quad E_{\underline{3}}=0\,,\cr
r<r_+\quad&& E_{\underline{1}}=-\frac{\mathcal{B} x}{\Big(1+\frac{\mathcal{B}^2}{4}  r^2 (1-x^2)\Big)^2}\,,\quad
B_{\underline{3}}=\frac{\mathcal{B} \sqrt{-f} \sqrt{1-x^2}}{\Big(1+\frac{\mathcal{B}^2}{4}  r^2 (1-x^2)\Big)^2}\,,\\
\hbox{PO}:\quad r>r_+\quad&& E_{\underline{1}}=-\mathcal{B}x\Big[2\mu\partial_r\Omega+\frac{(1+\mathcal{B}^2\mu rx^2)(1-\mathcal{B}^2\mu^2)}{\Omega}\Big]\,,\cr
&& E_{\underline{2}}=\mathcal{B}\frac{\sqrt{1-x^2}\sqrt{PQ}}{r\Omega}\,,\quad E_{\underline{3}}=0\,,\cr
r<r_+\quad&& E_{\underline{1}}=-\mathcal{B}x\Big[2\mu\partial_r\Omega+\frac{(1+\mathcal{B}^2\mu rx^2)(1-\mathcal{B}^2\mu^2)}{\Omega}\Big]\,,\quad
 B_{\underline{3}}=\mathcal{B}\frac{\sqrt{1-x^2}\sqrt{-PQ}}{r\Omega}\,.\nn
\eea
We follow the same Cartesian frame as \eqref{electric field in XZ} and plot the electric field lines in Fig.~\ref{Sch electric}. We find that the configurations of the SM and PO solutions are similar. The main difference is that the latter vacuum configuration has uniform color, indicating that the electric field strength is uniform. Neither solution possesses a cosmological horizon, and both differ significantly from the Fig.~\ref{LH electric} in the main text.

\end{document}